\documentclass[useAMS,usenatbib]{mn2e}

\usepackage[T1]{fontenc}
\usepackage{aecompl}

\usepackage{graphicx}
\usepackage[utf8]{inputenc}
\usepackage{amsmath}
\usepackage{amssymb}
\usepackage{color}
\usepackage{xspace}
\usepackage{tabularx}
\usepackage{multirow}
\usepackage{pbox}

\newcommand{\beq}{\begin{equation}}
\newcommand{\eeq}{\end{equation}}
\newcommand{\bey}{\begin{eqnarray}}
\newcommand{\eey}{\end{eqnarray}}

\newcommand{\Msun}{\ensuremath{\text{M}_\odot}}
\newcommand{\Lsun}{\ensuremath{\text{L}_\odot}}

\newcommand{\PhiN}{\phi}

\renewcommand{\d}{{\text{d}}}
\newcommand{\grad}{{\bf \nabla}}

\newcommand{\unit}[1]{\ensuremath{\,\text{#1}}}
\newcommand{\textforobjectname}[1]{\ensuremath{\,\text{\ \ \ \ [for #1]}}}

\newcommand{\mnras}{MNRAS}
\newcommand{\na}{NA}
\newcommand{\apj}{ApJ}
\newcommand{\apjl}{ApJL}
\newcommand{\apjs}{ApJS}
\newcommand{\aap}{A\&A}
\newcommand{\nat}{Nature}
\newcommand{\pasa}{PASA}

\newcommand{\prd}{Phys. Rev. D}

\newcommand{\aj}{AJ}
\newcommand{\araa}{ARA\&A}
\newcommand{\aapr}{A\&AR}
\newcommand{\ramses}{{\sc ramses}\xspace}

\title
[Expected properties of classical MW dSphs in Milgromian dynamics]
{A census of the expected properties of classical Milky Way dwarfs in Milgromian dynamics}

\author[F. L\"ughausen et al.]
{F.~L\"ughausen$^1$\thanks{fabian@astro.uni-bonn.de},
B.~Famaey$^2$
and P.~Kroupa$^1$
\\
\\
$^1$Helmholtz-Institut f\"ur Strahlen- und Kernphysik, Universit\"at Bonn, Nussallee 14--16, D-53115 Bonn, Germany\\
$^2$Observatoire astronomique de Strasbourg, Universit\'e de Strasbourg, CNRS, UMR 7550, 11 rue de l'Universit\'e, F-67000 Strasbourg, France
}

\begin{document}

\date{\today}
\maketitle

\begin{abstract}
Prompted by the recent successful predictions of the internal dynamics of Andromeda's satellite galaxies, we revisit the classical Milky Way dwarf spheroidal satellites Draco, Sculptor, Sextans, Carina, and Fornax in the framework of Milgromian dynamics (MOND). We use for the first time a Poisson solver with adaptive mesh refinement (AMR) in order to account simultaneously for the gravitational influence of the Milky Way and its satellites. This allows us to rigorously model the important external field effect (EFE) of Milgromian dynamics, which can reduce the effective acceleration significantly.
We make predictions on the dynamical mass-to-light ratio ($M_\text{dyn}/L$) expected to be measured by an observer who assumes Newtonian dynamics to be valid. We show that Milgromian dynamics predicts typical $M_\text{dyn}/L \approx 10$--$50 \,\Msun/\Lsun$.
The results for the most luminous ones, Fornax and Sculptor, agree well with available velocity dispersion data. 
Moreover, the central power-law slopes of the dynamical masses agrees exceedingly well with values inferred observationally from velocity dispersion measurements.
The results for Sextans, Carina and Draco are low compared to usually quoted observational estimates, as already pointed out by Angus. 
For Milgromian dynamics to survive further observational tests in these objects, one would thus need that either (a) previous observational findings based on velocity dispersion measurements have overestimated the dynamical mass due to, e.g., binaries and contaminant outliers, (b) the satellites are not in virial equilibrium due to the Milky Way tidal field, or (c) the specific theory used here does not describe the EFE correctly (e.g., the EFE could be practically negligible in some other theories), or a combination of (a)--(c). 
\end{abstract}


\maketitle

\section{Introduction}
\label{sect:intro}

Data on large scale structures, when interpreted in terms of Einstein's field equations, point towards a Universe dominated by dark energy and dark matter. Dark energy is generally represented by a cosmological constant, $\Lambda$, and dark matter (DM) is most often assumed to be made of hitherto undetected massive elementary particles, the so-called cold dark matter (CDM). Models based on less massive particles, so-called warm dark matter (WDM) lead largely to the same results, apart from some mild differences in the minimum mass of DM haloes and the presence of small constant density cores at their centre \citep{Maccio12}. However, at galaxy scales, the observations are in disagreement with many predictions based on particle DM \citep[e.g.][]{Kroupa1,Kroupa2}, whilst the observation of a tight correlation between the distribution of baryonic and missing mass seems to indicate that the effective law of gravity is well-described by Milgromian dynamics particularly in rotationally supported galaxies (\citealp{Mil83}, see \citealp{FamMcgaugh} for a major review, and also \citealp{Hernandez2014}, \citealp{Trippe2014}), rather than Newtonian dynamics plus DM.

The specific observed dynamics of spiral galaxies can be interpreted as becoming scale-invariant under transformations $(t,\boldsymbol x) \rightarrow (\lambda t,\lambda\boldsymbol x)$ with $\lambda \in \mathbb{R}$ when the accelerations fall well below the threshold acceleration $a_0 \approx 10^{-10}\,\mathrm{m}\,\mathrm{s}^{-2} \approx \Lambda^{1/2}$. 
This is mostly equivalent to stating that, in spherical symmetry, the gravitational attraction then approximately approaches $(g_\text{N} a_0)^{1/2}$, where $g_\text{N}$  is the classical Newtonian gravitational acceleration due to the baryonic matter.
This prescription, known as Milgromian dynamics, leads to a large body of remarkable predictions in galaxies \citep{FamMcgaugh}.
A general consequence of such dynamics is that, unlike Newtonian dynamics, it is nonlinear even in the nonrelativistic regime, meaning that it cannot satisfy the strong equivalence principle. For example, in the case of a satellite galaxy orbiting a more massive host galaxy, the satellite's internal dynamics is not independent from the acceleration it feels due to the external field of the host galaxy. The effect of this external acceleration on the internal dynamics of a system is known as the external field effect (EFE), and is very different from the tidal effect. For objects such as satellite galaxies, rigorously taking into account the EFE requires to account simultaneously for the gravitational influence of the host and the internal gravitational field of the satellites. 
In this work, we revisit the dynamics of dwarf spheroidal satellites of the Milky Way (MW) by making use of an advanced Poisson solver with adaptative mesh refinement (AMR).

Such dwarf spheroidal galaxies orbiting around more massive hosts range from $10^3$ to $10^7\,\Lsun$ with half-light radii of about $500\unit{pc}$ to $1\unit{kpc}$.
Two kinds of dwarf galaxies must exist in the framework of the standard cosmological model \citep[][and the references therein]{Kroupa2}: primordial dwarf galaxies (PDGs) and tidal dwarf galaxies (TDGs). 
PDGs formed early in the universe and are supposed to be embedded in small CDM haloes. Cosmological simulations have shown that a large number of PDGs as massive as $10^8\,\Msun$ and more should have formed as satellites orbiting the MW \citep{Klypin99,Moore99}. These primodial galaxies do not have preferred orbits and are thus roughly spherically distributed around the host, or only moderately flattened \citep{Wang2012}, and move in arbitrary directions. Even accretion from cold filaments has been demonstrated to not yield significant anisotropies \citep{Pawlowski2012b}.
TDGs on the other hand are dwarf galaxies resulting from major encounters of galaxies. In such encounters, gas and stars are stripped off the galaxies through tidal forces and form large tidal debris tails within which dwarf galaxies can form. Contrary to PDGs, TDGs can have only little or no cold or WDM \citep{Barnes92,Bournaud2010} and are clearly correlated in phase-space if they originate from the same event. They typically form vast disc-like structures around their past-encounter hosts.

Because the found dwarf spheroidal galaxies around the MW are observed to have extraordinary high dynamical mass-to-light ratios \citep[e.g.,][]{Mateo1991,Strigari2008,Walker2009,Walker2013,Battaglia2013}, they are generally thought to be PDGs enclosed in CDM subhaloes \citep[e.g.,][]{Belokurov2013}.
There are, however, a number of problems with this interpretation. The oldest one is known as the missing satellite problem: while there should be more than 500 nearly isotropically distributed CDM subhaloes with bound masses of $\gtrsim 10^8\,\Msun$ with a tidally limited size of $\gtrsim 1\unit{kpc}$ \citep{Moore99}, only 11 bright satellites have been detected (and only about 26 are known in total). It has been subsequently assumed that gas had collapsed to form substantial stellar populations only in some `lucky' CDM subhaloes, whilst the others would have lost their baryons or had stellar formation quenched for a variety of reasons \citep[e.g.][]{Brooks2013}, ranging from stellar feedback to tidal forces and reionization. Nevertheless, even in semi-analytical models taking such effects into account, there remain problems at the low-mass and high-mass end \citep[e.g.,][]{Kroupa1}. For instance, the most massive subhaloes of the MW in CDM simulations are too dense to host any of its bright satellites \citep[this is known as the `too big to fail' problem;][]{2big2fail}, leaving as a mystery why these massive haloes failed to form galaxies. 

Moreover, a second and even more problematic observation is that the dwarf spheroidal satellite galaxies of the MW are arranged in a corotating, vast polar structure \citep[VPOS,][]{VPOS}, which is completely incompatible with the predictions from CDM simulations. The same problem arises in the Andromeda galaxy \citep{Ibata2013Andromeda} where half of the satellites are rotating in an extremely thin planar structure oriented towards the MW.

The strong phase-space correlation of the satellites suggests that the observed satellites are not PDGs but TDGs. While this conclusion seems natural, it is in contradiction with CDM, because the dwarf satellites of the MW are observed to have very high dynamical mass-to-light ratios. The observations by \citet{Bournaud2007} also emphasize this conflict around external galaxies: they observe currently forming TDGs in the tidal debris of a galactic encounter, and these TDGs also possess a large amount of missing mass. This missing mass can, in the standard picture, only be explained by large amounts of unseen, presumably cold, molecular gas. The flat rotation curves of these dwarfs on the other hand are inconsistent with this expedient as they would require this baryonic DM to be distributed in an isothermal fashion. On the contrary, these rotation curves are well explained by Milgromian dynamics without any free parameters \citep{Gentile2007}.

So, if the conclusion that the MW dSphs are of tidal origin is true, the observed high dynamical mass-to-light ratios would imply that these objects are either out of equilibrium \citep{Kroupa1997,KlessenKroupa1998,Casas2012} or that a modified gravity scheme, such as those based on Milgromian dynamics, applies, or both. In the latter case, only those galaxies that appear to be in dynamical equilibrium should be compared to the static predictions of Milgromian dynamics \citep{McGaughWolf}.\footnote{While the faintest dwarf spheroidals show clear sign of being out of equilibrium, this is not the case for the most massive ones.} In the view of Milgromian dynamics, the tidal scenario seems very natural. Timing arguments suggest that M31 and the MW must have had a close tidal encounter, likely 7--11\unit{Gyr} ago \citep{timing}. In the standard model, this simple tidal encounter scenario is not possible at all, because the dynamical friction between the CDM haloes of the two encountering galaxies would lead to a galactic merger. The formation of TDGs around the MW could however be explained by other scenarios, e.g. the one modelled by \citet{Hammer2013arxiv}, but it is still in contradiction to the observed high amount of missing matter in these objects.

Since apparent high dynamical masses\footnote{The dynamical mass is the mass derived from the measured velocity dispersion under certain assumptions, e.g. dynamical equilibrium, while applying Newtonian dynamics.} (deduced when using classical Newtonian dynamics) are a natural property of Milgromian dynamics for objects of low surface density \citep{FamMcgaugh}, it is thus of high interest to predict what should be expected for the MW dwarf satellites in this context. This was pioneered for the MW dwarf spheroidals by \citet{Milgrom1995}, \citet{BradaMilgrom2000}, \citet{Angus08} and \citet{Hernandez2010}, while predictions for the Andromeda dwarfs were made by \citet{McGMil1} and \citet{McGMil2}. Correct a priori predictions were for instance made for the velocity dispersions of AndXVII, AndXIX, AndXX, AndXXI, AndXXIII, AndXXV, AndXXVIII. Among these, some are seen as outliers from the mass--luminosity--radius relations within the CDM paradigm because of their large size and low velocity dispersions, for instance AndXIX, AndXXI and AndXXV. On the contrary, these low velocity dispersions were correctly predicted a priori in Milgromian dynamics thanks to the EFE \citep{McGMil2}.

All these studies had the drawback of having to treat the EFE of Milgromian dynamics in a non-self-consistent manner. The external field indeed has a major influence on the predicted effective dynamical mass and has to be taken into account very carefully. This has recently been done properly in the work of \citet{Angus14} but without AMR, not allowing as much flexibility to study the various effects on vastly different scales. Here, we take advantage of the Milgromian Poisson solver with AMR, which we developed in the course of a larger project, in order to account simultaneously for the gravitational influence of the MW and its satellites. As a first application, we thus revisit the predictions for the brightest MW dwarfs, making predictions on the objects' dynamical mass-to-light ratios ($M_\text{dyn}/L$) expected to be measured when assuming Newtonian dynamics to be valid.

\section{Milgromian dynamics}\label{sect:mond}
Milgrom's simple formula, that is
\begin{equation}
	g = \left( g_\mathrm{N} a_0 \right)^{1/2}
\end{equation}
if $g_\mathrm{N} \ll a_0$ \citep{Mil83},
as such cannot be a final theory of gravity (e.g., no conservation of momentum). This formula arises from the approach of scale-invariance symmetry of the equations of motion under transformation $(t,r) \rightarrow (\lambda t, \lambda r)$ with $\lambda \in \mathbb{R}$ \citep{Milgrom2009}, and applies to spherically symmetric systems only. One can however derive theories of gravity that yield Milgrom's formula in spherical symmetry.
To date, many different generally covariant modified gravity theories reproducing Milgromian dynamics have been developed \citep{TeVeS,GEA,BIMOND}, and even at the classical level, various modified Poisson equations exist. While they could slightly differ out of spherical symmetry \citep{ZhaoFamaey}, the general predictions for dwarf spheroidal galaxies should be similar in all of these. One recent formulation \citep{QUMOND} has the following Poisson equation:
\begin{equation}
\nabla^2 \Phi=4 \mathrm{\pi} G \rho_\text{b} + \nabla \cdot \left[ \nu\left(|{\grad} \PhiN|/a_0\right) {\grad} \PhiN \right],
\label{eq:poisson}
\end{equation}
where $\rho_\text{b}$ is the baryonic density, 
$\Phi$ is the total (Milgromian) potential,
$\PhiN$ is the Newtonian potential such that $\nabla^2 \PhiN=4 \mathrm{\pi} G \rho_\text{b}$, and $\nu(x) \rightarrow 0$ for $x \gg 1$ and $\nu(x) \rightarrow x^{-1/2}$ for $x \ll 1$.
A family of functions fulfilling this definition of $\nu(x)$ \citep[see, e.g.][]{FamMcgaugh} is
\begin{equation}
\nu(x)=\left[{1+(1+4x^{-n})^{1/2} / 2}   \right]^{1/n} -1 .
\label{eq:nu}
\end{equation}
In the following, when not stated otherwise, we use the $n=1$ function, which is known to reproduce well the rotation curves of most spiral galaxies.

The second term in equation~(\ref{eq:poisson}) can be thought to be the matter density distribution $\rho_\text{ph}$ that would, in Newtonian gravity, yield the Milgromian boost to gravity, and is known as the ``phantom dark matter'' (PDM) density:
\begin{equation}
	\rho_\text{ph} = \frac{\nabla \cdot \left[ \nu\left(|{\grad} \PhiN|/a_0\right) {\grad} \PhiN \right]}{4\mathrm{\pi} G} \,.
\label{rho_phantom}
\end{equation}
This PDM density is not a real physical object but is only a mathematical description that allows us to solve the Poisson equation for Milgromian dynamics with only linear differential equations and one simple, algebraic step. In the framework of Newtonian dynamics, this mathematical source of gravity would be interpreted as missing matter or DM.

\subsection{Computing the effective dynamical masses predicted by Milgromian dynamics}
The PDM density that would source the Milgromian force field in Newtonian gravity
 is defined by equation~(\ref{rho_phantom}) and can be computed from the known classical (Newtonian) potential $\PhiN(\boldsymbol x)$.
To evaluate this term on a Cartesian grid, we make use of the grid-based scheme that has been devised by \citet{PRG}.

In order to treat the host and the satellite galaxies simultaneously, we implemented this scheme into the \ramses code \citep{Teyssier2002}: in this work, we make use of its Poisson solver \citep[see][for a detailed description]{GuilletTeyssier2011} and the available AMR infrastructure \citep[AMR,][]{Kravtsov97,Teyssier2002} to compute the effective Milgromian potential from the given distribution of baryonic matter. Starting from a coarse Cartesian grid, the AMR technique allows us to refine this grid on a cell-by-cell and level-by-level basis in the regions of interest: each cell which exceeds a given particle density (or equivalently barynonic mass density) is split into $2^3$ sub cells. This way, the potential of a large physical box containing structures of very different mass densities and sizes can be computed efficiently at a single time. In this work, we make use of this benefit and start with a bounding box that has a size that is large enough to host a Galaxy model at the centre as well as the satellite galaxies at their known positions. At the centre of this box, we place a three-dimensional mass density model of the MW determined by \citet{McGaugh2008_proceeding}. At the box boundaries, we use the Dirichlet boundary conditions
$\phi(r) = G M_\text{b} / r$
with $\phi(r)$ being the Newtonian potential at the distance $r$ to the centre of mass (of the whole baryonic density grid), and total baryonic mass $M_\text{b}$, to solve the Poisson equation for the Newtonian potential $\phi_N(\boldsymbol x)$. From this discrete potential, we compute the PDM density (equation~\ref{rho_phantom}) using the prescription from \citet[][their equation~4]{PRG}.

To find the gravitational potential $\Phi(\boldsymbol x)$ predicted by Milgromian dynamics, the resulting PDM density, $\rho_\text{ph}(\boldsymbol x)$, is added to the baryonic mass density, $\rho_\text{b}(\boldsymbol x)$. Poisson's equation, 
$\nabla^2 \Phi(\boldsymbol x) = 4\mathrm{\pi} G \left( \rho_\text{b}(\boldsymbol x) + \rho_\text{ph}(\boldsymbol x) \right)$,
now with the total effective dynamical mass (baryonic matter + PDM), is solved a second time, now using the boundary condition \citep[see Eq.~20 in][]{FamMcgaugh}
\beq
	\Phi(r) = \left(G M_\text{b} a_0\right)^{1/2} \ln (r)
\eeq
on the last grid point at the distance $r$ to the centre of mass of the \textit{baryonic} density grid with total baryonic mass $M_\text{b}$.

In this work, we consider only static models of the MW satellites and use the PDM density to predict the effective dynamical mass\footnote{We prefer to use the term `dynamical mass-to-light ratio' rather than classically just `mass-to-light ratio', because in the Milgromian picture, PDM is not real matter but a mathematical construction. Dynamical mass however makes clear that we refer to the baryonic mass plus the DM equivalent as deduced by a Newtonian observer in a Milgromian universe.} of these satellite galaxies.
The grid is resolved to a resolution of 10\,pc and less (the typical half-mass radius of the considered dSphs is of the order of $500-1000\unit{pc}$). The resolution limit is visible in the logarithmic plots at small radii, $r$.

\subsection{External field effect (EFE)}\label{sect:efe}
Contrary to classical, linear, Newtonian dynamics, Milgromian dynamics is described by a non-linear theory and breaks the strong equivalence principle. 
As a consequence, the internal dynamics of a satellite system (e.g. dSphs around the MW, or galaxies in the external field of a galaxy cluster) does not decouple from the external field produced by its mother system.
This means, the external field can drastically reduce the `acceleration boosting effect' of Milgromian dynamics with respect to classical dynamics, it can even break it completely down to Newtonian behaviour if the external acceleration (and thus the ``total" acceleration) is larger than $a_0$.
If $g_\text{N,ext}$ is the external (Newtonian) acceleration, and $g_\text{N,int}$ the internal one, then the EFE does not play a role if 
$$g_\text{N,ext} \ll g_\text{N,int} \,.$$
In this case, the system is in the Newtonian or Milgromian regime, and its dynamics depends only on $g_\text{N,int}$.\footnote{However, even in this regime, the external field induces a small quadrupole which can, e.g., be measured with high-precision experiments in the Solar system \citep{MilgromSolar,Hees14}.}

On the other hand, if the acceleration due to the external field dominates and is in the Newtonian regime, 
$$g_\text{N,int} < a_0 \ll g_\text{N,ext} \,,$$
the external acceleration field takes over and the internal dynamics appear purely Newtonian even when $g_\text{N,int} < a_0$. 
In between these extreme cases, i.e. if the external field dominates but is itself well below $a_0$,
$$g_\text{N,int} < g_\text{N,ext} < a_0 \,,$$ 
the system is then Newtonian with a renormalized gravitational constant \citep{FamMcgaugh}.
That is, in the framework of QUMOND, the spatial distribution of the PDM density is proportional to that of the baryons, $\rho_\text{ph}(\boldsymbol x) \propto \rho_\text{b}(\boldsymbol x)$.
This is referred to as the quasi-Newtonian regime.
The exact behaviour of the EFE finally depends on the particular theory, and particularly on the applied $\nu$-function.
This peculiar property of any theory implementing Milgromian dynamics is in contrast to the experience of our classical Newtonian thinking and challenges our intuition.

The EFE has another remarkable consequence that concerns the cusp/core problem \citep{CuspCoreProblem}: 
while simulations show that haloes made of CDM have cuspy profiles with a central matter density profile, $\rho(r) \propto r^\alpha$, that has a power-law slope of approximately $\alpha_\text{NFW} = \left.\d\log\rho_\text{NFW}(r)/\d \log r\right|_{r\rightarrow 0} = -1$, the profiles of the effective dynamical mass (observationally inferred from measured velocity dispersions) of observed dSphs appear to be cored\footnote{In the literature, the term ``cuspy" is commonly used if the powerlaw slope is steeper (i.e. less) than the inner slope of the NFW profile, $\alpha <-1$, while ``cored" refers to $\alpha > -1$.} \citep[e.g.,][]{Walker2011}, i.e. $\alpha = \left.\d\log\rho_\text{DM/ph}(r)/\d\log r\right|_{r\rightarrow 0} \approx 0$.
In Milgromian dynamics, the power-law slope is expected to be $\approx -0.5$ if the model is isolated (i.e. without external field) and if the baryonic matter density distribution itself is cored \citep{Milgrom2009d}. The EFE can however reduce this slope from $\approx -0.5$ to $\approx 0$ if the galaxy's dynamics are dominated by the external field, because in this case the PDM profile has the same shape as the baryonic matter profile with a different normalization constant (see above), i.e. $\alpha = \left.\d\log\rho_\text{ph}(r)/\d\log r\right|_{r\rightarrow 0} \approx \left.\d\log\rho_\text{b}(r)/\d\log r\right|_{r\rightarrow 0} \approx 0$. We discuss in the results section how this applies to each object individually.
In Section~\ref{sect:varyD} and Fig.~\ref{fig:varyD}, we demonstrate how an example model of a MW satellite is affected by the gravitational potential of its host galaxy.

In the course of this work, the EFE is self-consistently implemented, because the computed models contain the MW and its satellite all at one time.

\section{Models}
\subsection{Dwarf spheroidal models}

Dwarf spheroidals are quasi-spherical galaxies, the observed density profiles are mostly well fitted by King density models \citep{King66}, 
\beq\label{eq:king}
	\rho(r) \propto \frac
		{ \arccos(z)/z - \sqrt{1 - z^2} }
		{ \mathrm{\pi} r_\text{king} \left[ 1 + \left(r_\text{lim}/r_\text{king}\right)^2 \right]^{3/2} z^2 }
\eeq 
with $z^2 = \left( 1 + r/r_\text{king} \right)^2 / \left( 1 + r_\text{lim}/r_\text{king} \right)^2$ \citep{King62}.
King model fits have been performed for the classical dSphs e.g. by \citet{Strigari2008}, for reviews see \citet{FergusonBinggeli1994} and \citet{Mateo1998}. 
We adopt the King models and scale the given luminosities and size parameters to the recent distances compiled by \citet{McConnachie12}. 
The model parameters are listed in Table~\ref{tab:models}, their positions in Galactocentric coordinates in Table~\ref{tab:positions}.

\begin{figure*}
\centering
\includegraphics[width=8.5cm]{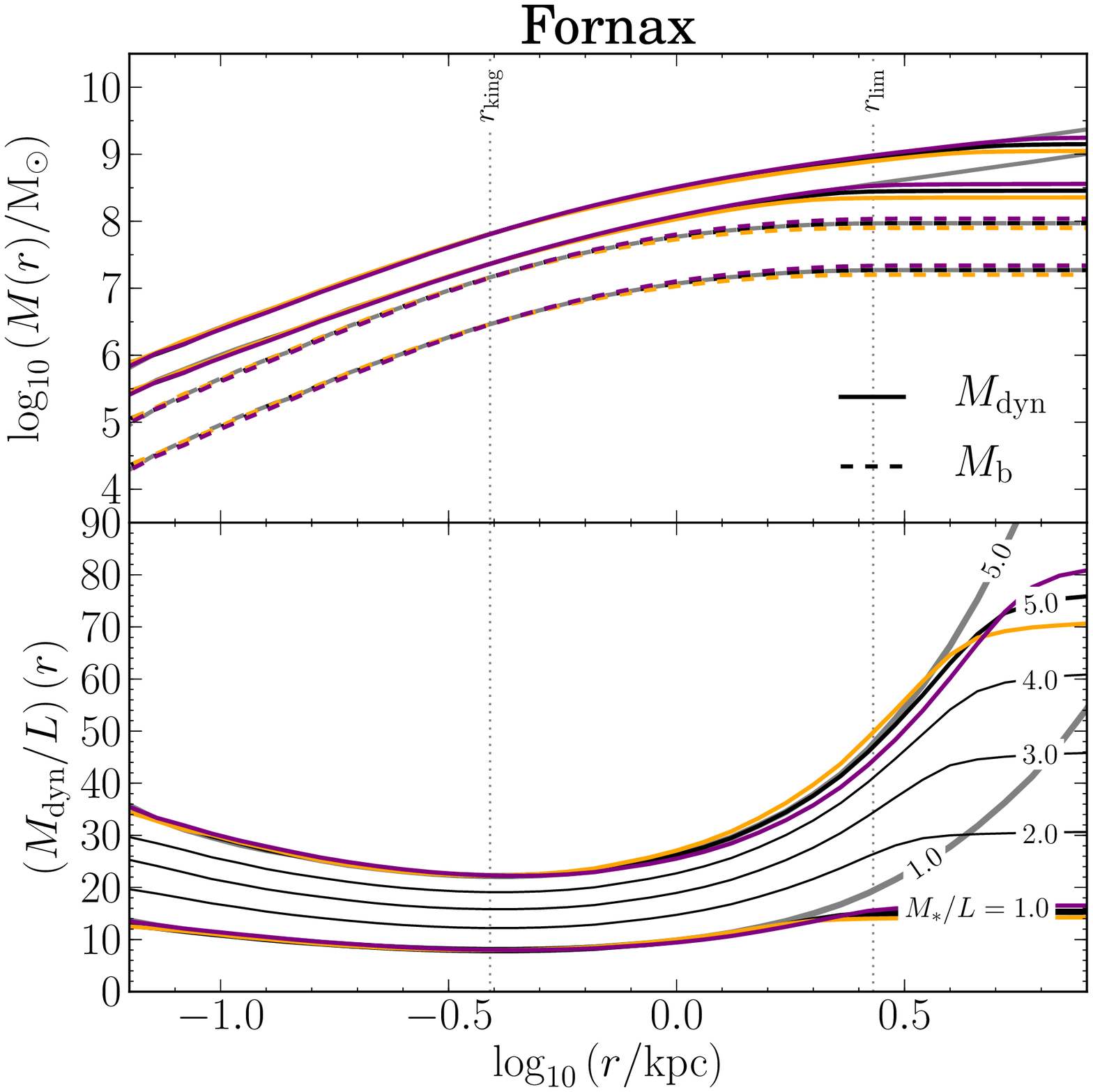}
\caption{
	Results for Fornax.
	Top panel: 
		cumulative profiles of the assumed baryonic mass (dashed line) and the predicted effective dynamical mass (solid line) as a function of distance from Fornax's centre. The corresponding mass model paramters are detailed in Table~\ref{tab:models}.
		The lower dashed lines belong to the $M_*/L=1\,\Msun/\Lsun$ model,
		the upper dashed lines to the $M_*/L=5\,\Msun/\Lsun$ one.
	Bottom panel: 
		the ratio of effective dynamical mass (baryonic matter + PDM) to baryonic mass with radius $r$ are plotted for each Fornax model.
		The results are discussed in Section~\ref{sect:fornax}.
	Colour coding:
		the black lines correspond to models at their most likely (``normal") distances ($D$, see Table~\ref{tab:positions}). Purple lines represent the models at the maximum distances to the Sun, $D_\text{max}$, and orange lines at $D_\text{min}$, respectively. Grey lines show the models in isolation, i.e. without external field. All lines are marked with the individual values of $M_*/L$ in units of $\Msun/\Lsun$.
		The values of $r_\text{king}$ and $r_\text{lim}$ are given for the normal distances, $D$.
}
\label{fig:fornax}
\end{figure*}

\begin{figure*}
\centering
\includegraphics[width=8.5cm]{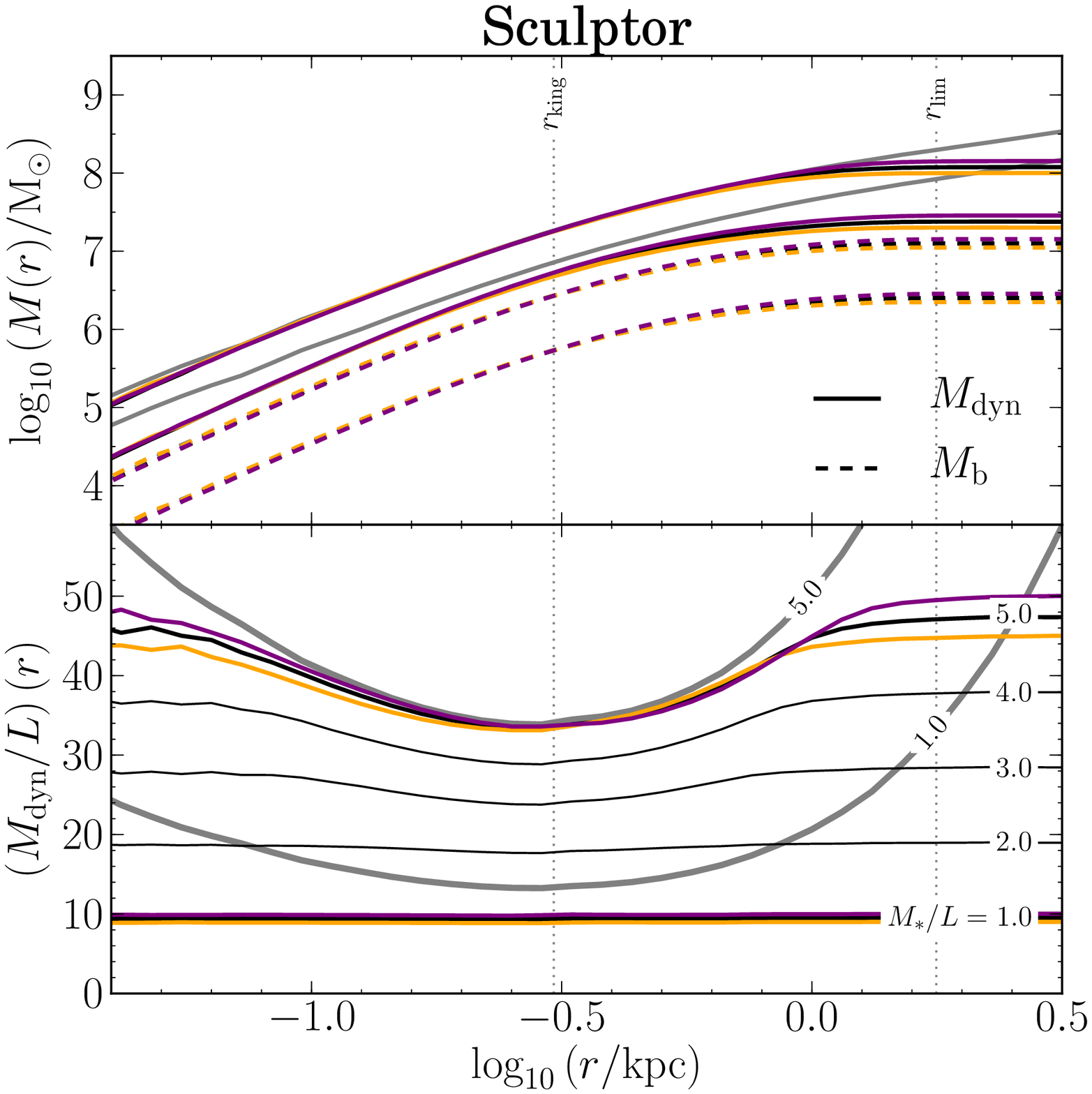}
\caption{
	Results for Sculptor. 
	See the caption of Fig.~\ref{fig:fornax} for an explanation.
	The results for this object are discussed in Section~\ref{sect:sculptor}.
	}
\label{fig:sculptor}
\end{figure*}

\begin{figure*}
\centering
\includegraphics[width=8.5cm]{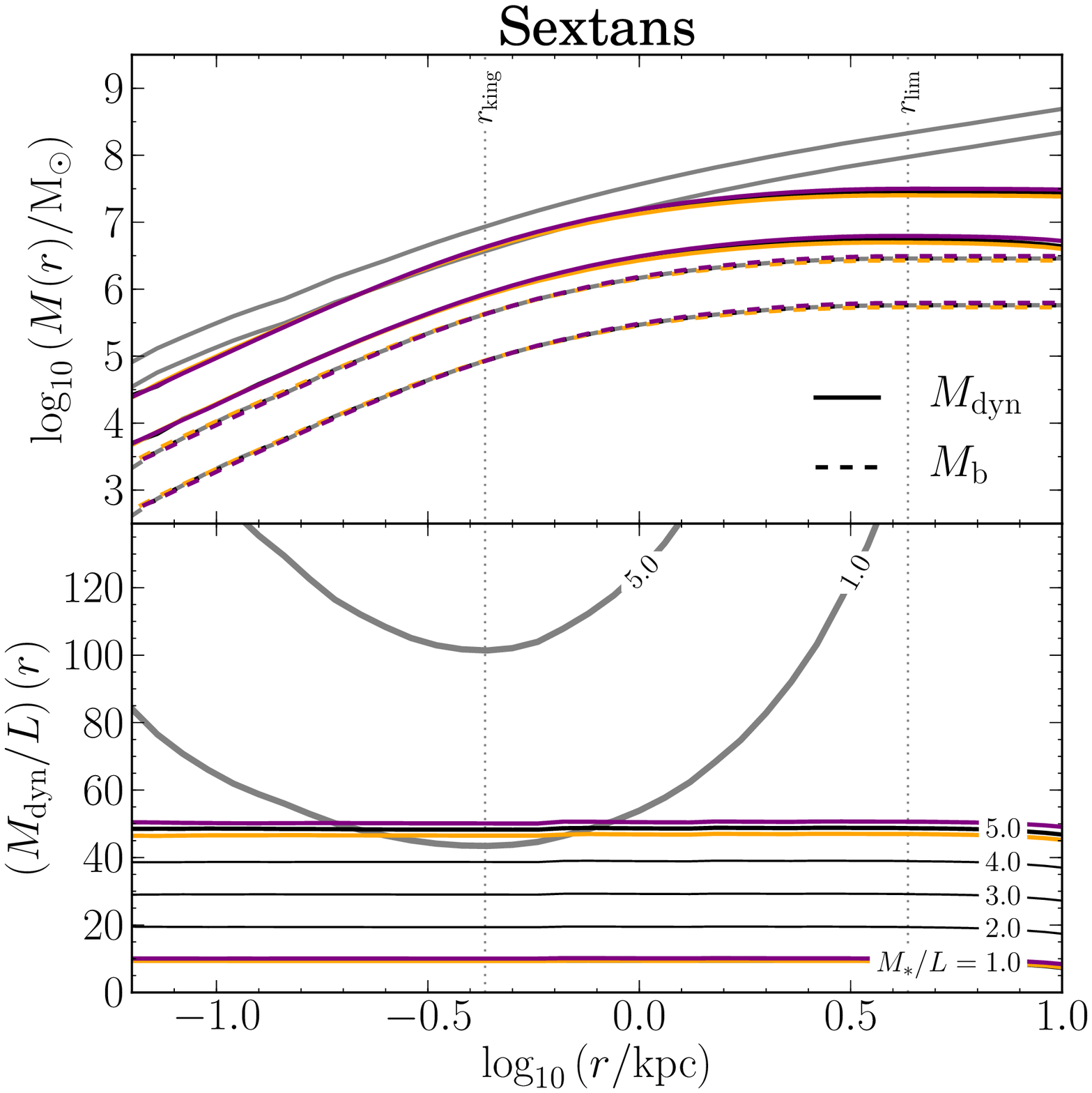}
\caption{
	Results for Sextans.
	See the caption of Fig.~\ref{fig:fornax} for an explanation.
	The results for this object are discussed in Section~\ref{sect:sextans}.
}
\label{fig:sextans}
\end{figure*}

\begin{figure*}
\centering
\includegraphics[width=8.5cm]{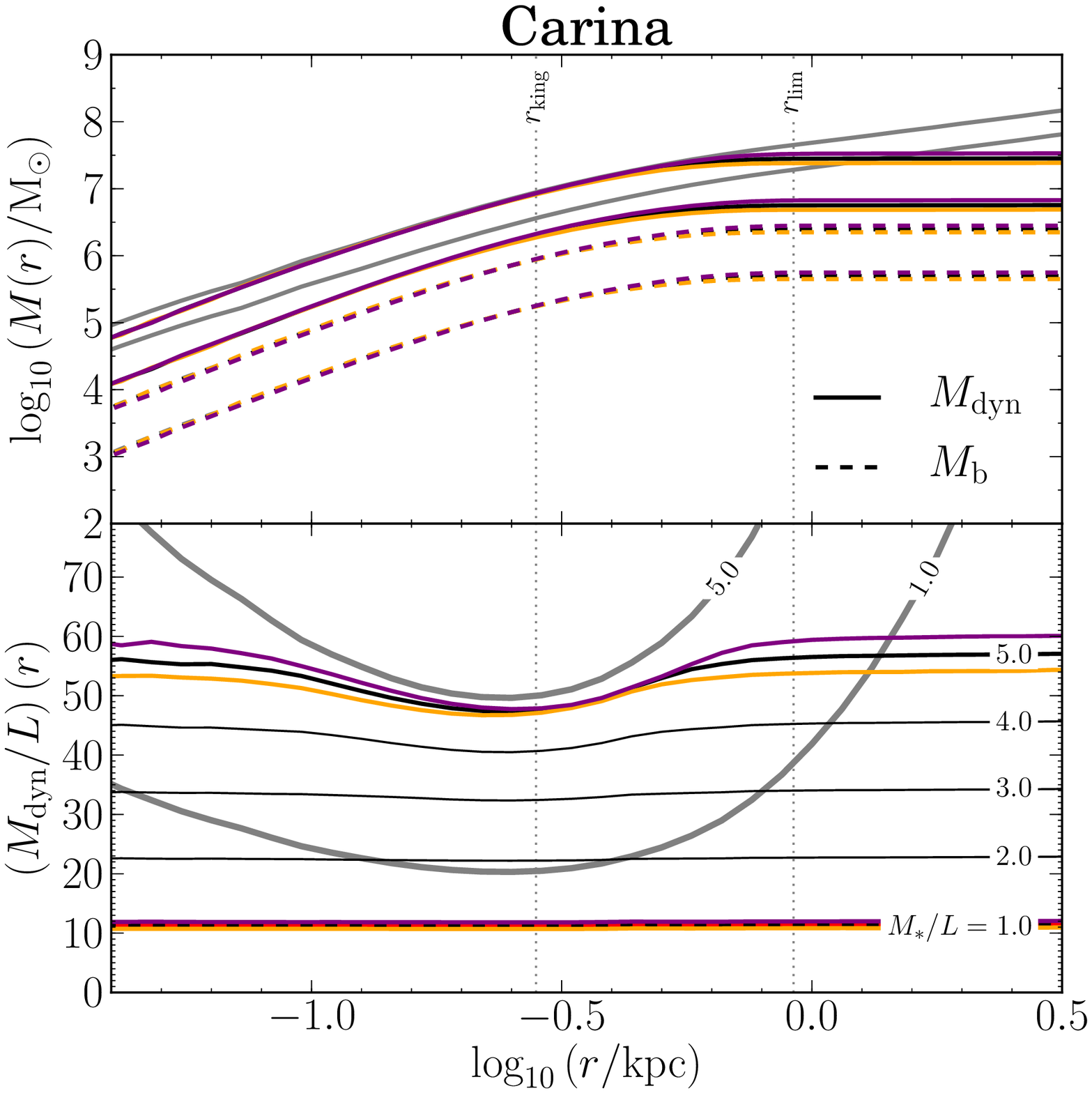}
\caption{
	Results for Carina.
	See the caption of Fig.~\ref{fig:fornax} for an explanation.
	The results for this object are discussed in Section~\ref{sect:carina}.
}
\label{fig:carina}
\end{figure*}

\begin{figure*}
\centering
\includegraphics[width=8.5cm]{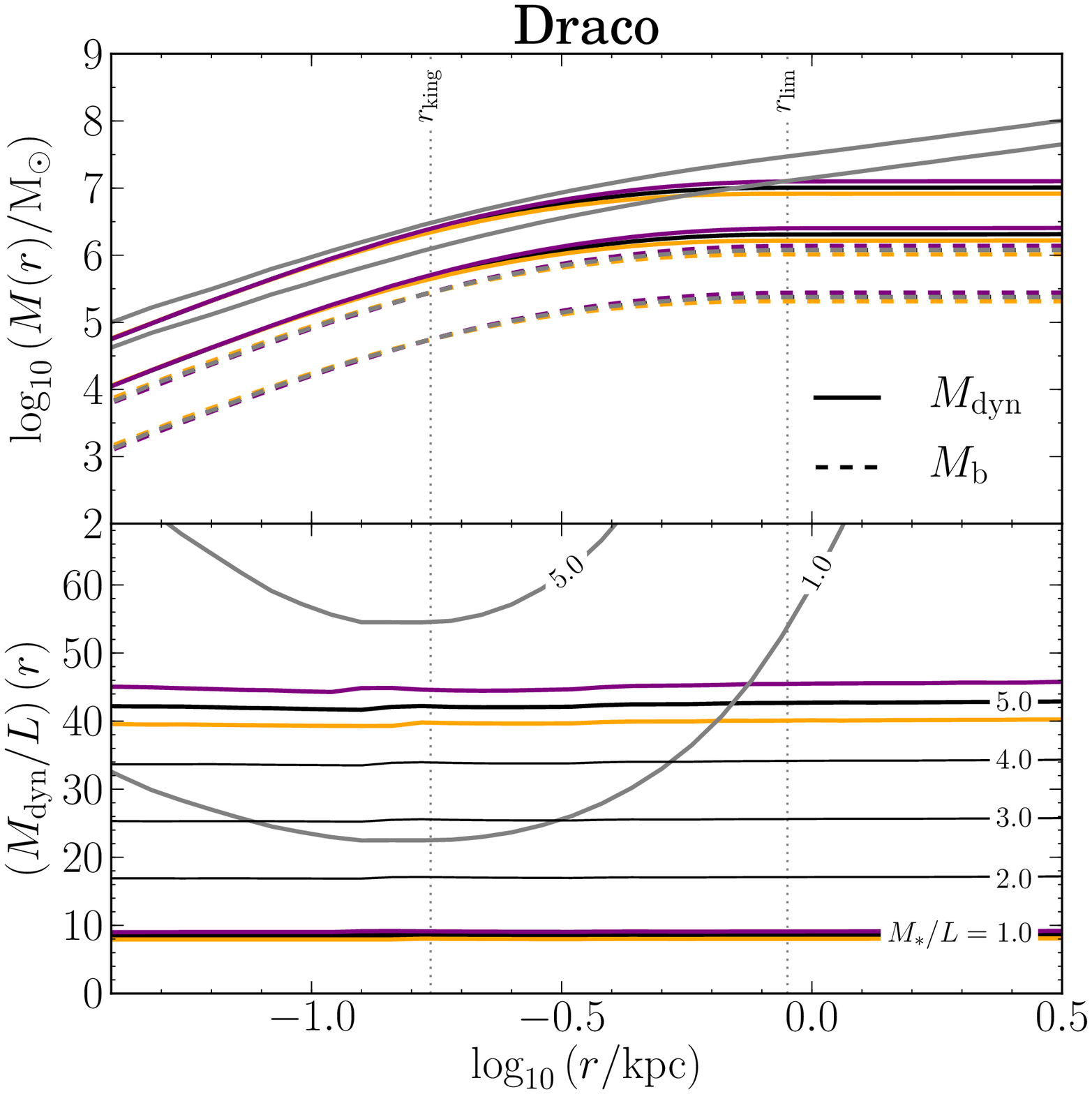}
\caption{
	Results for Draco.
	See the caption of Fig.~\ref{fig:fornax} for an explanation.
	The results for this object are discussed in Section~\ref{sect:draco}
	}
\label{fig:draco}
\end{figure*}

In this work, we consider the so-called classical (luminous) dwarf spheroidal galaxies, ordered by decreasing total luminosity.
We do not include Ursa Minor (UMi), because it appears to be out of equilibrium \citep{Kleyna2004}.
For each object, we investigate the following 11 models. 
\begin{itemize}
\item At the most likely distance to the Sun \citep[given by][]{McConnachie12}, we provide models with $M_*/L=1,2,3,4,\text{ and }5\,\Msun/\Lsun$. These models are plotted with black solid lines (Figs~\ref{fig:fornax}--\ref{fig:draco}).
\item At the minimum and maximum distance (given by the 1-$\sigma$ errors), we provide models with $M_*/L=1\text{ and }5\,\Msun/\Lsun$ to demonstrate how the distance error transfers to the results. These models are plotted with purple ($D_\text{max}$) and orange ($D_\text{min}$) solid lines (Figs~\ref{fig:fornax}--\ref{fig:draco}).
\item We further provide isolated models without external field with $M_*/L=1\text{ and }5\,\Msun/\Lsun$ (grey solid lines in Figs~\ref{fig:fornax}--\ref{fig:draco}). For each galaxy, the isolated model with $M_*/L=5\,\Msun/\Lsun$ represents the upper limit of $M_\text{dyn}/L$ that can be achieved with this implementation of Milgromian dynamics.\footnote{Under the aforementioned assumptions/simplifications like spherical symmetry and dynamical equilibrium.}
\end{itemize}

\subsection{MW model}
We use one of the MW mass model from \citet{McGaugh2008}.
The model features a stellar exponential disc with scale length of $R_\mathrm{d} = 2.3\,\mathrm{kpc}$, scaleheight of 0.3\,kpc, and a total mass of $4.9\times10^{10}\,\Msun$. Moreover, it has a thin gaseous disc of $1.2\times10^{10}\,\Msun$ and a bulge made of a Plummer model with $0.6\times10^{10}\,\Msun$ and a half-mass radius of 1\,kpc. 

Although the MW potential is modelled in much detail, this is not crucial, because the spatial size of the satellites is much smaller than the size of the MW disc and their Galactocentric distances (which are 80\,kpc and more). More important are the total masses of the individual Galactic components.

In the Galactocentric coordinate system we use, the Sun is located at the $x$-axis at $8.5\unit{kpc}$. All specified Galactocentric coordinates are given with respect to this system. The positions of the satellite galaxy models are varied along their line of sight as seen from the position of the Sun.

\begin{table*}
\parbox{0.65\textwidth}{\caption{
	List of all King models used. The model parameters (King radius, limiting radius and luminosity) are adopted from \citet{Strigari2008} and are scaled appropriately for the considered distances $D$, $D_\text{min}$ and $D_\text{max}$, where $D$ is the most likely distance (between the object and the Sun) and $D_\text{min}$ and $D_\text{max}$ are the minimum and the maximum distances (1-$\sigma$ deviation of $D$ along the line of sight).
	The distances are adopted from the compilation of \citet{McConnachie12}.
	The respective Galactocentric distances and positions are provided in Table~\ref{tab:positions}.
}}
\begin{tabularx}{0.65\textwidth}{lllllll}
	\hline
	\\[-1.5ex]
	& 
	$r_\text{king}/\text{kpc}$ & 
	$r_\text{lim}/\text{kpc}$ & 
	$L_V/(10^5\,\Lsun)$ &
	$D/\text{kpc}$ &
	$D_\text{min}/\text{kpc}$ & 
	$D_\text{max}/\text{kpc}$ \\
	\\[-1.5ex]
	\hline
	\\[-1.5ex]
	Fornax  & 0.429 & 2.972 & 188 & 151.9 & 140.3 & 164.6 \\
	Sculptor  & 0.305 & 1.773 & 25.4 & 87.0 & 81.8 & 92.6 \\
	Sextans  & 0.432 & 4.321 & 5.8 & 92.9 & 81.1 & 96.9 \\
	Carina  & 0.281 & 0.919 & 5.0 & 109.2 & 103.2 & 115.7 \\
	Draco  & 0.173 & 0.894 & 2.4 & 76.9 & 71.3 & 82.9 \\
	\\[-1.5ex]
	\hline
\end{tabularx}
\label{tab:models}
\end{table*}

\begin{table}
\parbox{0.475\textwidth}{\caption{List of all positions in Galactic coordinates (MW centre at $[0,0,0]$, Sun at $[8.5\,\text{kpc},0,0]$) from \citet{McConnachie12}. $R$ is the according distance to the Galactic Centre. The satellite positions are varied along the line of sight within the $1\sigma$ measurement errors \citep[see also][]{McConnachie12}. $R_\text{min}$ and $R_\text{max}$ are the distances to the Galactic Centre at the positions closest to and farthest away from the Sun respectively.
}}
\begin{tabularx}{0.475\textwidth}{lllll}
	\hline 
	\\[-1.5ex]
	& 
	$\text{Position } [x,y,z]$ & 
	$R$ &
	$R_\text{min}$ & 
	$R_\text{max}$ \\
	& / kpc & / kpc & / kpc & / kpc \\
	\\[-1.5ex]
	\hline 
	\\[-1.5ex]
	Fornax & $-41.5, -51.0, -134.1$ & 149.4 & 162.0 & 137.7 \\
	Sculptor & $-5.4, -9.8, -85.3$ & 86.0 & 91.7 & 80.7\\
	Sextans & $-36.9, -56.9, 57.8$ & 89.1 & 93.1 & 85.2  \\
	Carina & $-25.2, -95.9, -39.8$ & 106.8 & 113.4 & 100.8 \\
	Draco & $-4.6, 62.2, 43.2$ & 75.9 & 70.2 & 82.0 \\
	\\[-1.5ex]
	\hline
\end{tabularx}
\label{tab:positions}
\end{table}

\section{Discussion of the model variables}\label{sect:discussvariables}
The used dSph models are determined by (i) the density model and its parameters, (ii) the total luminosity, (iii) the stellar mass-to-light ratio\footnote{In the more general context, ``stellar mass-to-light ratio" should actually mean ``baryonic mass"-to-``stellar light" ratio, because it relates the stellar luminosity to the baryonic mass of an object, which includes not only stellar mass, but all kind of baryonic matter. In the case of the dSphs, this is the same, because there is essentially no gas.} and (iv) the position with respect to the Galactic Centre which specifies the external gravitational field felt by the satellite galaxy. In the literature, a large number of different models with different parameters can be found for each of the dSphs. In the following subsections, we start with toy models based on the Sculptor model (see Section~\ref{sect:sculptor}) to investigate and discuss in which way the available variables affect the dynamical mass-to-light ratio, $M_\text{dyn}/L$, as expected by the applied formulation of Milgromian dynamics.
Remember that $M_\text{dyn}/L$ is the dynamical mass-to-light ratio deduced from the dynamics of the stars by an observer when using Newtonian dynamics.
This approach is intended to provide the reader a basic understanding of how the model parameters (particularly the external field and the stellar mass-to-light ratio) affect the PDM halo in the Milgromian picture of galactic dynamics.

\subsection{The external field}\label{sect:varyD}
\begin{figure}
\centering
\includegraphics[width=8.5cm]{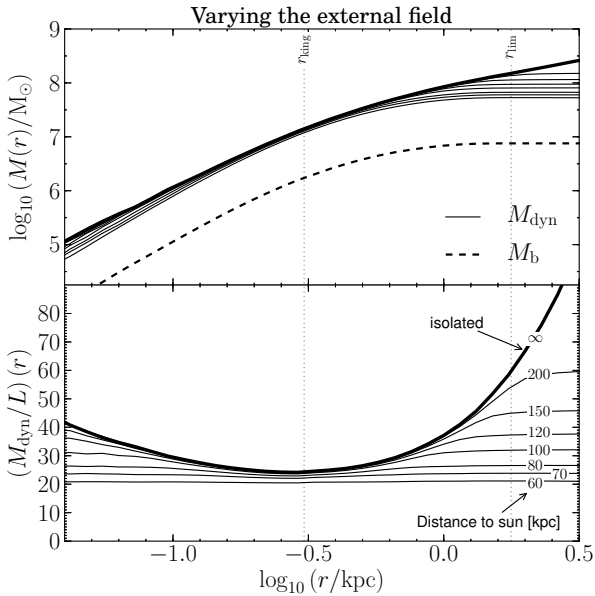}
\caption{
	The lines show the cumulative mass profiles of the same King model (based on Sculptor with $M_*/L=3\,\Msun/\Lsun$, $D=87\unit{kpc}$) at different Galacticocentric distance to illustrate the effect of the external field (see Section~\ref{sect:varyD}).
	Top panel: the cumulative mass (the assumed baryonic mass and the predicted dynamical mass) enclosed within the radius $r$ is shown.
	Bottom panel: the ratio of the effective dynamical mass (phantom DM + baryonic matter) to the baryonic matter content are presented as a function of distance $r$ to the centre of the dwarf galaxy.
	The thick solid line shows the model in isolation, i.e. without external Galactic field. 
	For the detailed description see Section~\ref{sect:varyD}.
	}
\label{fig:varyD}
\end{figure}
Because Milgromian dynamics is, by virtue of its scale-invariant property, acceleration-based, the external gravitational field plays always a prominent role (see Section~\ref{sect:efe}). Fig.~\ref{fig:varyD} demonstrates how the external field affects the shape of the PDM halo of a sample dSph model. The thick solid lines shows the Sculptor model (with $D=87\unit{kpc}$, $M_*/L=3\,\Msun/\Lsun$; see Section~\ref{sect:sculptor}) in isolation or without external field, i.e. at infinite Galactocentric distance. If the distance of this satellite model to the centre of the Galactic potential is decreased, i.e. the strength of the external gravitational field is increased, the total mass of its PDM halo decreases (i.e. the effect of Milgromian dynamics weakens). This is because the overall acceleration in the satellite galaxy is enhanced by the external field, which affects particularly the central and outer regions (where the internal Newtonian acceleration, $g_\text{N}$, is low) because of the non-linearity of $\nu(x)$. As the external field increases, it ``cuts off" the PDM density in the lowest-internal-acceleration parts (see the upper panel of Fig.~\ref{fig:varyD}), making the PDM density follow the baryonic density, until the dynamical mass-to-light ratio appears nearly constant at all $r$ (see the lower panel of Fig.~\ref{fig:varyD}).
In the latter case, the external field of the MW dominates the satellite's internal dynamics ($g_\text{N,int} < g_\text{N,ext} < a_0$). Consequently, in this case the satellite's effective dynamical mass profile follows its baryonic mass profile, $\rho_\text{dyn}(r) \propto \rho_\text{ph}(r) \propto \rho_\text{b}(r)$.
Moreover, this means that, if the shape of the baryonic matter density is cored, also the PDM halo has to be cored in this external field-dominated case (cf. Section~\ref{sect:efe}).

As the internal + external acceleration approaches the limit $g_\text{N,int}+g_\text{N,ext} \gg a_0$, the dynamical mass to light ratio approaches the stellar one, $M_\text{dyn} \rightarrow M_*$.

\subsection{Stellar mass-to-light ratio, $M_*/L$}
\label{sect:varyML}
\begin{figure}
\centering
\includegraphics[width=8.5cm]{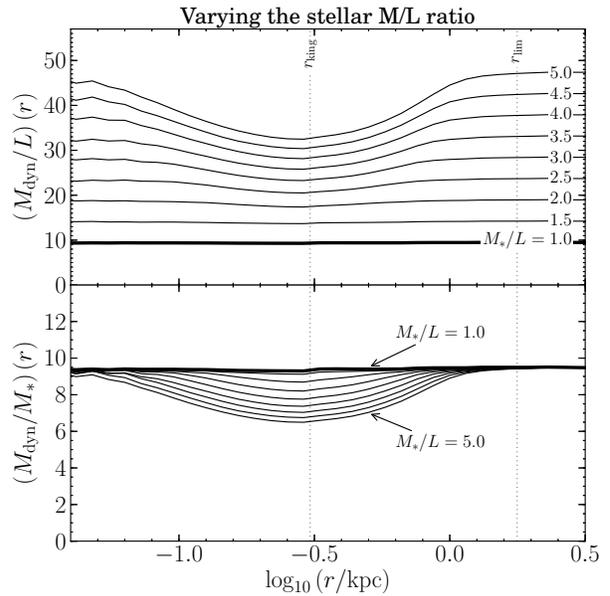}
\caption{
	Each line shows the same model (based on Sculptor) with different stellar mass-to-light ratios, $M_*/L$ (the total stellar luminosity, $L$, is kept constant, $D=87\unit{kpc}$). This sequence of models demonstrates the impact of the $M_*/L$ uncertainty on the $M_\text{dyn}/L$ ratio predicted by Milgromian dynamics.
	Top panel: the cumulative mass (the assumed baryonic mass and the predicted dynamical mass) enclosed within the radius $r$ is shown.
	Bottom panel: the ratio of the effective dynamical mass (phantom DM + baryonic matter) to the baryonic matter content are presented as a function of distance $r$ to the centre of the dwarf galaxy.
	The thick solid line shows the model in isolation, i.e. without external Galactic field. 
	For the detailed description see Section~\ref{sect:varyML}.
}
\label{fig:varyML}
\end{figure}
The total mass of a stellar system is commonly inferred from the total luminosity, $L$, by knowing the or assuming a reasonable stellar mass-to-light ratio, $M_*/L$. In most cases, the value of this variable is not well constrained and left as a fit parameter. The dynamical mass-to-light ratio prediction (within the framework of Milgromian dynamics viewed by a Newtonian observer) however is very sensitive to this quantity, because the dynamical mass density distribution is computed from the baryonic density distribution. We evaluate all models with $M_*/L=1$ and $5\,\Msun/\Lsun$ to provide lower and upper limits on $M_\text{dyn}/L$.

Again based on the Sculptor model, Fig.~\ref{fig:varyML} shows how the predicted dynamical mass-to-light ratio, $M_\text{dyn}/L$, changes with $M_*/L$, where $1, \dots, 5\,\Msun/\Lsun$ is a reasonable range for the stellar mass-to-light ratio for these dwarf spheroidals.

If $M_*/L = 1\,\Msun/\Lsun$, the satellite galaxy is dominated by the external field of the MW but is still in the deep Milgromian regime ($g_\text{N,int} < g_\text{N,ext} \ll a_0$, cf. Section~\ref{sect:efe}),\footnote{In this particular model, this is the case if $M_*/L \lesssim 2$.} and the satellite's effective dynamical mass profile thus follows its baryonic mass profile, so that the dynamical mass-to-light ratio, $M_\text{dyn}/L$, is constant at all radii $r$, and $M_\text{dyn}/L \propto M_*/L$.

If $M_*/L$ (and thus the baryonic density and accordingly $g_\text{N,int}$) is increased ($M_*/L \gtrsim 2$), the external field becomes less dominant, $g_\text{N,ext} < g_\text{N,int} < a_0$, while staying in the deep Milgromian regime. The strength of the EFE is decreased and the observable dynamical mass-to-light ratio (observationally inferable from velocity dispersion measurements) would then become radius-dependant (see Fig.~\ref{fig:varyML}).

\subsection{Total luminosity, $L$}\label{sect:varyL}
\begin{figure}
\centering
\includegraphics[width=8.5cm]{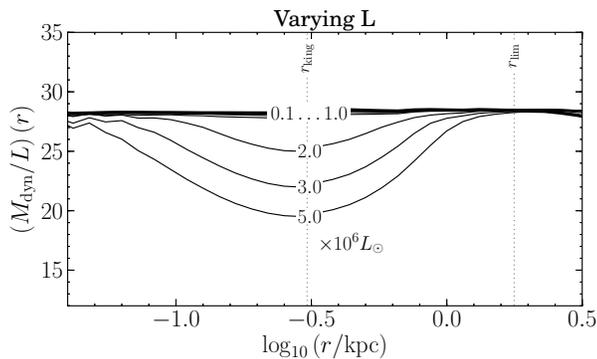}
\caption{
	The ratio of the predicted effective dynamical mass to baryonic matter is plotted as a function of radius. The shown models are based on Sculptor ($M/L=3\,\Msun/\Lsun$, $D=87\unit{kpc}$), the total luminosity, $L$, is varied.
	See also the description in Section~\ref{sect:varyL}.
}
\label{fig:varyL}
\end{figure}
Varying the total luminosity, $L$, means varying the total mass (because the stellar mass-to-light ratio is kept constant) and thus varying the average density (because the size parameters are kept constant as well) and accordingly $g_\text{N,int}$ varies. Fig.~\ref{fig:varyL} shows the dynamical mass-to-light ratio of the Sculptor model for different $L$. As long as the total mass is small such that the internal dynamics is dominated by the external field, the dynamical mass-to-light ratio predicted by Milgromian dynamics does not depend on $L$. But as the baryonic mass increases and the internal accelerations get to the order of $a_0$, this degeneracy vanishes and $M_\text{dyn}/L$ becomes radius-dependant.

\subsection{Density model and its radial parameters}\label{sect:varyR}
\begin{figure}
\centering
\includegraphics[width=8.5cm]{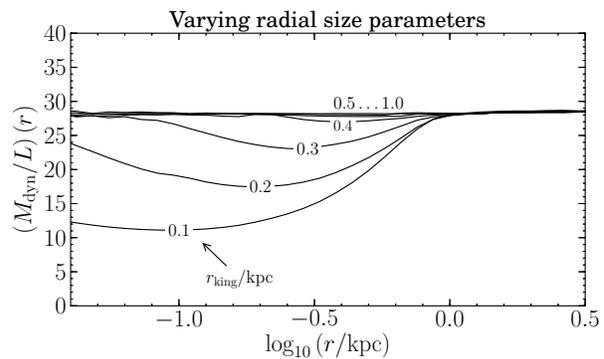}
\caption{
	The ratio of the predicted effective dynamical mass to baryonic matter is plotted as a function of radius. The presented models are based on Sculptor ($M/L=3\,\Msun/\Lsun$, $D=87\unit{kpc}$), the radial model paramters (King radius and limiting radius) are variied.
The King radius, $r_\text{king}$, is located at the local minimum of $M_\text{dyn}/L$.
	See also the description in Section~\ref{sect:varyR}.
}
\label{fig:varyR}
\end{figure}
For each dSph galaxy, a number of different models and fits can be found in the literature. These are Plummer and King models, exponential profiles, power-law profiles and also the profile derived by \citet{Zhao97}. In most cases, the truncated King model provides the best fit to the observed luminosity profiles. 

To demonstrate visually the influence of the external field on the dynamical mass-to-light ratio (see Fig.~\ref{fig:varyR}),
we vary, as before based on the Sculptor model, the radial size paramters from $r_\text{king}=0.1$ to $5\,\text{kpc}$ while setting the limiting radius to $r_\text{lim} = 5.3 \, r_\text{king}$. The actual King radius is $\approx 0.3\unit{kpc}$. The total luminosity and the stellar mass-to-light ratio are kept constant.  Fig.~\ref{fig:varyR} shows the resulting $M_\text{dyn}/L$ ratios. 

If the King radius is large (in this particular model $\approx 0.5$--$1\unit{kpc}$), the system has a low density, therefore low internal accelerations $g_\text{N}$, and the internal dynamics are dominated by the external field. In this case, the total effective mass profile follows the baryonic mass profile and $M_\text{dyn}/L$ is independent of the exact radial density model.

If however the baryonic matter density becomes compact/dense enough such that the internal accelerations, $g_\text{N,int}$, become large enough to leave the external field-dominated regime (here $r_\text{king} < 0.5\unit{kpc}$), the density model becomes important and sensitively affects the dynamical mass-to-light ratio at $r\ll r_\text{lim}$.

\bgroup
\renewcommand\arraystretch{1.2}
\begin{table*}
\parbox{0.85\textwidth}{\caption{
	A compilation of dynamical mass-to-light ratios from the literature are listed (in units of $\Msun/\Lsun$) side by side with the predictions made in this paper applying Milgromian dynamics.
	$M_{0.1}$ and $M_{0.3}$ are the total dynamical masses within 0.1 and $0.3\unit{kpc}$ from the dSph's centre, as found by \citet{Strigari2008} from velocity dispersion measurements when applying Newtonian dynamics.
	These masses are divided by the respective luminosities, $L_{\mathrm{V},0.1}$ and $L_{\mathrm{V},0.3}$, provided by the same authors.
	$M_{r_\mathrm{max}}$ is the total dynamical mass within the limiting radius as provided by \citet{Walker2008} (see Table~\ref{tab:models}).
	The total $V$-band luminosity, $L_{V,\mathrm{tot}}$, is again adopted from \citet{Strigari2008}.
}}
\begin{tabularx}{0.85\textwidth}{lccccccc}
	\hline
	\\[-2.5ex]
	& 
	\multicolumn{2}{c}{ $M_{0.1}/L_{V,0.1}$ } &
	\multicolumn{2}{c}{ $M_{0.3}/L_{V,0.3}$ } &
	\multicolumn{3}{c}{ $M_{r_\mathrm{max}}/L_{V\mathrm{tot}}$ }\\
	& 
	\multirow{2}{1.5cm}{\centering predicted with EFE}  &  
	\multirow{2}{1.5cm}{\centering observ. inferred}  &  
	\multirow{2}{1.5cm}{\centering predicted with EFE}  &  
	\multirow{2}{1.5cm}{\centering observ. inferred}  &  
	\multirow{2}{1.5cm}{\centering predicted with EFE}  &  
	\multirow{2}{1.8cm}{\centering predicted without EFE}  &  
	\multirow{2}{1.5cm}{\centering observ. inferred} \\
	\\
	\\[-2.5ex]
	\hline
	\\[-2.5ex]
        Fornax &  $[10.9,29.9]$ & $12.9^{+7.5}_{-4.3}$ &  $[8.1,22.8]$ & $6.8^{+0.5}_{-0.7}$ &  $[14.3,47.9]$ &  $[21.2, 51.6]$  & 12 \\	
	Sculptor  & $[8.9,40.5]$ & $40^{+74}_{-26}$ &  $[8.9,33.7]$ & $23^{+2}_{-7}$ &  $[8.9,50.1]$ &  $[33.3,78.7]$  & 38 \\
	Sextans  & $[9.5,50.3]$ & $280^{+93}_{-47}$ &  $[9.5,50.3]$ & $143^{+113}_{-35}$ &  $[9.5,50.3]$ &  $[163.8,370.6]$  & 108 \\
	Carina & $[10.7,54.5]$ & $293^{+43}_{-37}$ &  $[10.7,48.0]$ & $81^{+10}_{-5}$ &  $[10.7,59.4]$ &  $[38.6,90.5]$  & 81 \\
	Draco  & $[8.0,44.7]$ & $55^{+122}_{-12}$ &  $[8.0,44.7]$ & $137^{+15}_{-21}$ &  $[8.0,44.7]$ &  $[53.8, 125]$  & 346 \\
	\\[-2.5ex]
	\hline
\end{tabularx}
\label{tab:literature}
\end{table*}
\egroup

\section{Results}
Figs~\ref{fig:fornax}--\ref{fig:draco} present the results for the considered dSph satellite galaxies. 
The top panels show the cumulative mass profiles of the baryonic matter (dashed lines) and the resulting effective dynamical mass (i.e. mass of baryonic matter + PDM, solid lines).
The bottom panels show the dynamical mass-to-light ratio as function of the distance $r$ to the respective satellite galaxy's centre. The black lines represent models at the most likely distance, $D$ (see Table~\ref{tab:positions}). Purple lines show models at the maximum distance to the Sun, $D_\text{max}$, orange lines those at the minimum distance, $D_\text{min}$. All lines are marked with their model-specific values of $M_*/L$ in units of $\Msun/\Lsun$.

\subsection{Fornax}\label{sect:fornax}
Due to its large distance to the Galactic Centre and its large total luminosity (i.e. mass), 
Fornax (see Fig.~\ref{fig:fornax}) is effectively isolated and unaffected by the EFE.
The resulting dynamical mass-to-light ratio, which is presented in the lower panel of Fig.~\ref{fig:fornax}, depends on the radius, $r$, and ranges between 10 and $50\,\Msun/\Lsun$ depending on the model. Contrary to the external field-dominated dSphs, the accuracy of the model has a strong influence on the resulting $M_\mathrm{dyn}/L$ as a function of radius.

Because this dSph is effectively isolated and has a cored baryonc matter density distribution, 
the central shape of the PDM profile has a power-law slope of $\alpha = \left.\d\log\rho_\text{ph}(r)/\d\log r\right|_{r \rightarrow 0} \approx -0.5$.
This is consistent with the inference\footnote{\citet{Walker2011} derive the quantity $\Gamma = \alpha + 3$.} by \citet{Walker2011}, who find $\alpha = -0.39^{+0.43}_{-0.37}$.
Although the external field has almost no influence on Fornax' internal dynamics, it truncates the PDM halo at $r \gtrsim r_\mathrm{lim}$.

Values of $M_\mathrm{dyn}/L$ found in the literature are remarkably small compared to those of other dSphs. Their observational errors cover a range from $6.1\,\Msun/\Lsun$ to only $20.4\,\Msun/\Lsun$. These values agree well with our results if $M_*/L \approx 1\,\Msun/\Lsun$.

\subsection{Sculptor}\label{sect:sculptor}
Sculptor's dynamics is on the verge to being in the external field-dominated regime. It is external field-dominated if $M_*/L = 1\,\Msun/\Lsun$, but it is clearly not if $M_*/L = 5\,\Msun/\Lsun$ (see Fig.~\ref{fig:sculptor}, and also the discussion of model parameters in Section~\ref{sect:discussvariables} based on Sculptor).
Fig.~\ref{fig:sculptor} illustrates that $M_\text{dyn}/L \approx 9.5 \, M_*/L$
if $M_*/L \lesssim 2$. 
For larger stellar mass-to-light ratios, $M_*/L > 2$, we find that the EF becomes less dominant and the dynamical mass-to-light ratio becomes radius-dependant, most prominently at the core radius $r_\textrm{king}$.

In general, the relation 
$M_\text{dyn}/L \approx 9.5 \, M_*/L$ 
therefore holds true only at the limiting radius, $r_\text{lim}$.
In contrast to fully EFE-dominated galaxies, the results for Sculptor are sensitive not only to $M_*/L$ and $g_\text{N,ext}$, but also to the exact density model and total luminosity. One has to be careful with radial dependences. 
The central density profile of the PDM halo is cored or very close to be cored (i.e. $\alpha \approx 0$, depending on the exact model).
This slope is well consistent with the inference by \citet{Walker2011}, who find $\alpha = -0.05^{+0.51}_{-0.39}$.

The Sculptor results agree very well with the values found in the literature if $M_*/L \approx 4\,\Msun/\Lsun$. Notably, the literature values show the expected trend that the dynamical mass-to-light ratio is similar at small radii (0.1\,kpc) and large ($r_\textrm{max}$) radii, and lower in between (even though this is only a trend given the large observational errors).

\subsection{Sextans}\label{sect:sextans}
Sextans' distance to the Galactic Centre is not much larger than that of Scuptor, but in comparison it is much fainter (by a factor of $\approx 1/5$), bringing it into the external field-dominated regime:
The effective dynamical mass follows the baryonic mass and 
$$M_\text{dyn}/L \approx 9.7 \, M_*/L  \textforobjectname{Sextans}.$$
The choice of the exact density model and the exact total luminosity are of minor importance (within certain limits of course), and the stellar mass-to-light ratio as well as the strength of the external field are entirely determining the resulting effective dynamical mass-to-light ratio. Also Sextans' PDM density profile is clearly cored.

The dynamical mass-to-light ratios we derived for Sextans under the assumptions of Milgromian dynamics to describe gravity correctly are far below the values found by \citet{Walker2008} and \citet{Strigari2008}, whose errors cover the wide range from 108 to $373\,\Msun/\Lsun$. These values are even far above the computed upper Milgromian limit (if the satellite is assumed to be isolated).

\subsection{Carina}\label{sect:carina}
Although Carina is clearly more distanced from the Galactic Centre than Sculptor, the effect of the external field on the internal dynamics appears very similar, because in both galaxies the ratio of $g_\text{N,int}$ to $g_\text{N,ext}$ is similar. Also Carina appears clearly external field-dominated if $M_*/L =1\, \Msun/\Lsun$, but partly overcomes this effect if the actual $M_*/L$ is high (e.g.~$5\,\Msun/\Lsun$, as shown in Fig.~\ref{fig:carina}). 
The average dynamical mass-to-light ratio is relatively high compared to the other dSphs, it is 
$\approx 11.2 \, M_*/L$ 
if 
$M_*/L \lesssim 2.5 \,\Msun/\Lsun$.
The central shape of Carina's PDM halo is cored or \textit{very} close to be cored (closer than in the case of Sculptor), as can be seen in the bottom panel of Fig.~\ref{fig:carina}: $\left.\d (M_\text{dyn}/L)/ \d r \right|_{r \rightarrow 0} \approx 0$.

In the literature, we again find much larger values for $M_\mathrm{dyn}/L$. These cover the range from $336\,\Msun/\Lsun$ in the central region to $81\,\Msun/\Lsun$ in the outer region. The inferred values therefore indicate a cuspy DM profile, whereas we expect the PDM profile to be cored in theory. However, given the large scatter of these values which are inferred from observations, one should use them with much care, because the dynamical mass is estimated from the velocity dispersion, and large velocity dispersions can have various origins. Altogether, our determined values for $M_\mathrm{dyn}/L$ are however a bit too low compared to those inferred from observations \citep{Angus14}.

\subsection{Draco}\label{sect:draco}
Draco is the faintest of the classical satellites, and it is located at a small Galactic distance and is thus strongly influenced by the gravitational potential of the MW. As a consequence, its internal dynamics are dictated by this external field, and the dynamical mass-to-light ratio (see Fig.~\ref{fig:draco}) is thus radius independent: 
$\rho_\text{dyn}(r) \propto \rho_\text{b}(r)$, and 
$M_\text{dyn}/M_* \approx 8.5$.
Thus
$$M_\text{dyn}/L \approx 8.5 \, M_*/L \textforobjectname{Draco}$$
The central shape of the PDM halo is thus cored, because the baryonic density profile is cored.
Furthermore, the predicted $M_\text{dyn}/L$ are neither sensitive to the exact density model (see Section~\ref{sect:varyR}) nor to the observational uncertainty of Draco's total luminosity (see Section~\ref{sect:varyL}). What matters are the stellar mass-to-light ratio and the strength of the external field.

The dynamical mass-to-light ratios we found for Draco are very small compared to those available in the literature (cf. Table~\ref{tab:literature}). For example, 
\citet{Strigari2008} find 
$M_\mathrm{dyn}/L_\mathrm{V}=55^{+122}_{-12}\,\Msun/\Lsun$ within $r=0.1\unit{kpc}$, and
$137^{+15}_{-21}\,\Msun/\Lsun$ within $r=0.3\unit{kpc}$.\footnote{These values for the dynamical mass-to-visible light ratios are deduced from the density models and luminosities given by \citet{Strigari2008}, see their table~1.} \citet{Walker2008} find $346\,\Msun/\Lsun$ within $r=r_\mathrm{max}$.\footnote{We combine the dynamical mass measured by \citet{Walker2008} with the stellar luminosity given by \citet{Strigari2008}. The results are compiled in Table~\ref{tab:literature}.}
While the first value agrees fairly with our findings, the overall conclusion is that our results are in contradiction with either the values of $M_\mathrm{dyn}/L$ from velocity dispersion measurements (which could be contaminated by binaries and outliers) or with the assumption that Draco is in dynamical equilibrium, or both.

\section{Summary and conclusions}
The problem of the nature and dynamics of the dwarf spheroidal satellite galaxies is a vivid one. As highlighted in numerous recent studies, their phase-space distribution around the MW and the Andromeda galaxy is not compatible with them being primordial galaxies embedded in CDM haloes \citep[e.g.,][]{Ibata2014,Pawlowski14prep}. 
On the contrary, if they are of tidal origin, they can contain only little or no DM. In this case, the observed high velocity dispersions conclude either that {\it all} these objects must be out of equilibrium, or that Newtonian dynamics fails on this scale and that a different theory of gravity must apply (e.g., Milgromian dynamics).
In spiral galaxies, the correlation between the mass discrepancy and the gravitational acceleration has long been known to hold for orders of magnitude in mass, and can be interpreted as evidence for Milgromian dynamics. Such dynamics naturally predicts that the MW and Andromeda must have had a close tidal encounter, likely 7--11\unit{Gyr} ago \citep{timing}, leading to the formation of at least a significant fraction of today's satellites of the Local Group galaxies. Recent predictions of internal velocity dispersions of Andromeda's satellites within Milgromian dynamics have proven very successful \citep{McGMil1,McGMil2}. For the MW dwarfs, the situation is less clear. It has long been known that ultra-faint dwarfs cannot be accounted for in Milgromian dynamics if they are in dynamical equilibrium \citep{McGaughWolf}: these objects are close to fully filling their Milgromian tidal radii, and therefore are likely out of equilibrium. For classical dwarfs, we revisited the dynamics here (apart from UMi which also appears out of equilibrium), by taking advantage of the AMR Poisson solver to solve for the MW and the dwarf satellites simultaneously. We produced a table of predicted dynamical mass-to-light ratios which can be useful for observers (Table~\ref{tab:literature}).

We find typical $M_\text{dyn}/L$ of $\approx 8$ to $50\,\Msun/\Lsun$ (depending on model parameters, particularly the stellar mass-to-light ratio). In the case of Sculptor and Fornax, these values agree well with observations. In the case of Draco, Sextans, and Carina, these values are low compared to todays observational findings. This is in accordance with what Angus~(2008) had found, and it can mean that 
\begin{enumerate}
\item 
	the satellites are not in virial equilibrium due to the MW tidal and external field, 
\item 
	past observational findings are incorrect due to outliers and binary contamination, or 
\item 
	that the specific modified gravity theory used\footnote{Note also that we implemented here only one particular $\nu$-function} is not the theory that describes the EFE correctly. 
	For the latter case, we provide for each satellite upper limits of $M_\text{dyn}/L$ possible in Milgromian dynamics, in case the external field turns out to be negligible.
\end{enumerate}

It has already been argued in the past that the EFE might be an observational problem of Milgromian dynamics as formulated here, when confronting predictions to data \citep{Scarpa2006,Hernandez2010,Hernandez2012a,Hernandez2012b,Hernandez2013}. The argument is that, often, when the EFE starts playing a role, the agreement of Milgromian dynamics with observational data becomes marginal, while it remains good if the EFE is neglected: this might indeed be true for the dwarf spheroidals of the MW considered here. However, it is not necessarily the case in general. For instance, the escape speed from the MW can be determined from the EFE and agrees well with observations \citep{Famaey07}, and nearby open clusters having internal accelerations below $a_0$ do not exhibit large mass discrepancies. Also, in the CDM context, some dwarfs close to M31 have been pointed out as outliers because of their low velocity dispersions, while with Milgromian dynamics, such small velocity dispersions are naturally predicted \citep{McGMil2}: this prediction relies on the EFE being non-negligible as in this paper. Nevertheless, we should point out that, even though the EFE is a necessary consequence of Milgromian dynamics, in some implementations of the theory, it could be negligible in practice: this can be the case for instance in time-nonlocal modified inertia theories \citep{Milgrom2011}. Computations of, e.g., the escape speed from the MW would in this case become more complicated and many concepts such as the escape speed could have to be fully redefined. 
In view of the current inferences of dynamical masses of the MW dwarfs, this absence of EFE should certainly be kept as a possibility, as advocated in \citet{Scarpa2006,Hernandez2010,Hernandez2012b,Hernandez2013,Hernandez2012a}.

\citet{Kroupa1997} has shown that it is possible to achieve high $M_\text{dyn}/L \approx 100$ even in DM-free dSphs by assuming purely classical Newtonian dynamics.
The reason is that the satellites that were set up with spherical phase-space distribution functions evolve away from this state by losing particles from outer regions of phase-space due to the Galactic tides. The assumption made by the observer who assumes spherically symmetric equilibrium structures is then wrong, leading to very high apparent $M_\text{dyn}/L$ values, despite the models not having any DM.
This finding also applies to Milgromian dynamics (and of course also to PDGs embedded in CDM haloes), although we expect that the effect is less strong \citep{Hernandez2012a}.

That observational findings of the measured dynamical mass are not as correct as we think today is also one possibility which should not be excluded a priori. Dynamical masses are derived from the velocity dispersion, which is usually based on measurements that are very sensitive to effects that have not been taken into account yet, e.g. the number of binary stars, or plain outliers from the background. \citet{Serra10} have for instance shown that taking into account outliers was bringing Sextans back on the Milgromian prediction. A similar expectation can be made for Draco and Carina. 
Interestingly, it has recently been shown that the tidal effects are not significantly changing the predictions for Carina \citep{Angus14}.

We note that the predictions of Milgromian dynamics are most accurate for the most luminous satellites, and least for the less luminous ones.
The most luminous dwarf galaxies likely have had the highest star formation rates (SFRs) in the past. High SFRs result in high minimum embedded star cluster masses, making the embedded clusters denser and thus destroying binaries more efficiently \citep{Marks2011}. It is therefore likely that the velocity dispersion measurements of the less luminous dwarfs are more affected by unresolved binaries than those of the most massive dwarfs.
If one finds that the dynamical masses computed here (which are predictions of Milgromian dynamics based on static equilibrium models, and taking into account the EFE) are compatible with future, more precise measurements of the velocity dispersions in these dwarf galaxies, this would strengthen the notion that the MW dSph satellite galaxies are TDGs that have been formed $7-11\unit{Gyr}$ ago as a consequence of a close encounter between M31 and the MW. In this case, $N$-body computations based on Milgromian dynamics of the MW--M31 encounter should further test this tidal scenario.

If it however turns out that all measured velocity dispersions are correct and that the considered dwarf galaxies {\it are} in virial equilibrium, then the computed dynamical masses based on Milgromian dynamics tell us that the specific implementation used here can be excluded, and one has to consider other theories, such as modified inertia theories in which the EFE can be practically negligible \citep{Milgrom2011}.


\subsection*{Acknowledgements}
FL is supported by DFG grant \mbox{KR1635/16-1}. We thank the referee X. Hernandez for a careful reading of the manuscript and useful suggestions.



\end{document}